\documentclass{llncs}
\usepackage{amsmath,amsfonts}
\usepackage{algorithmic}
\usepackage{algorithm}
\usepackage{array}
\usepackage[caption=false,font=normalsize,labelfont=sf,textfont=sf]{subfig}
\usepackage{textcomp}
\usepackage{stfloats}
\usepackage{url}
\usepackage{xcolor}
\usepackage{verbatim}
\usepackage{graphicx}
\usepackage{xspace}
\usepackage{cite}
\newcommand{\paper}{paper\xspace}

\begin{document}

\title{A Desynchronization-Based Countermeasure Against Side-Channel Analysis of Neural Networks}


\author{%
Jakub Breier\inst{1,2}  \and 
Dirmanto Jap\inst{3} \and 
Xiaolu Hou\inst{4} \and 
Shivam Bhasin\inst{3} 
}%
\institute{
Silicon Austria Labs, TU-Graz SAL DES Lab, Graz, Austria
\and
Graz University of Technology, Austria
\\
email: jbreier@jbreier.com 
\and
Nanyang Technological University, Singapore \\ email: \{djap,sbhasin\}@ntu.edu.sg
\and
Slovak University of Technology, Bratislava, Slovakia
\\email:houxiaolu.email@gmail.com
}



\maketitle

\begin{abstract}
Model extraction attacks have been widely applied, which can normally  be used to recover confidential parameters of neural networks for multiple layers.
Recently, side-channel analysis of neural networks allows parameter extraction even for networks with several multiple deep layers with high effectiveness. 
It is therefore of interest to implement a certain level of protection against these attacks.

In this paper, we propose a desynchronization-based countermeasure that makes the timing analysis of activation functions harder.
We analyze the timing properties of several activation functions  and design the desynchronization in a way that the dependency on the input and the activation type is hidden.
We experimentally verify the effectiveness of the countermeasure on a 32-bit ARM Cortex-M4 microcontroller and employ a t-test to show the side-channel information leakage.
The overhead ultimately depends on the number of neurons in the fully-connected layer, for example, in the case of 4096 neurons in VGG-19, the overheads are between 2.8\% and 11\%.
\end{abstract}


\section{Introduction}
Current deep learning models grow to millions of parameters and are being widely deployed as a service.
As the training of such networks requires large amounts of data and computing time, the organizations that created the models tend to keep them proprietary.
Another motivation to keep the models confidential is to reduce the success rate of adversarial attacks that are more powerful in a white-box setting, i.e., when the attacker knows the model parameters.
This motivated the development of model extraction attacks on neural networks that either try to mimic the model behavior or extract the parameter values~\cite{lowd2005adversarial}.

The ``standard'' model extraction attacks try to get the proprietary information on the model by using a set of well-designed queries.
This, however, normally does not allow the full extraction of model parameters (also called \textit{exact extraction}), but rather a weaker type of extraction, which can be \textit{functionally equivalent extraction}, \textit{fidelity extraction}, or \textit{task accuracy extraction}~\cite{jagielski2020high}.
On the other hand, with the help of hardware attack vectors, which can be considered especially in the case of embedded machine learning models~\cite{batina2022implementation}, it is possible to extract the exact values of the parameters, either with side-channel analysis~\cite{batina2019csi,chmielewski2021reverse} or with fault injection attacks~\cite{breier2021sniff}.

In~\cite{batina2019csi}, it was shown that the information about the used activation function can be observed by measuring the time of the function computation.
The timing behavior is dependent on the type of the function and the input data and there are clear patterns forming after a certain range of inputs was fed to the activation function.
In this paper, we aim at removing those patterns to make the timing analysis useless for the attacker.
We utilize a desynchronization-based countermeasure that adds random delay to the function computation so that no matter what the activation type and the input data, the timing measurement gives random outputs.
We experimentally show that our method works by using a 32-bit ARM Cortex-M microcontroller as a device under test.

\section{Background}

\subsection{Side-Channel Analysis}

In~\cite{kocher1996timing}, it has been shown that even though cryptographic algorithms are proven to be theoretically secure, their physical implementation can leak information regarding confidential data, such as a secret key. This is commonly referred to as Side-Channel Analysis (SCA). An adversary could typically exploit different means of physical leakages, such as timing, power, electromagnetic (EM) emanation, etc. Thus, by observing these leakages, an adversary could analyze and deduce the secret information being processed, by utilizing statistical methods, such as correlation, etc.

\subsection{Side-Channel Countermeasures}

Since the idea of SCA is that the leakage could provide information regarding the internally processed data, the goal of protection techniques is to minimize or remove these dependencies.
Different countermeasures have been proposed to protect against SCA, and overall, these can be classified into \textit{hiding} and \textit{masking} techniques. 

The hiding countermeasures aim at breaking the relationship between the processed data and the leakage, for example, using desynchronization, shuffling, etc. The general idea is to introduce noises in the measurement to make the attacks harder.

On the other hand, for masking countermeasures, the aim is to remove the relation by introducing randomness to mask the actual data being processed. In this case, the data being processed will be masked with different random mask value for every execution and without the knowledge of the random mask, the attacker could not recover the actual intermediate value.

\subsection{Hardware Attacks on Neural Networks}

Recently, SCA has been applied to attacking neural network implementations. The aim is to recover the secret model, which might be sensitive, for example, if the model has been trained with confidential data or if the model is a commercial IP that has been trained and subjected to piracy. Hence, determining the layout of the network with trained weights is a desirable target for an attacker. Thus, using SCA, the attacker could reverse engineer the neural networks of interest by using some additional information that becomes available while the device under attack is operating. 

Several attacks have been reported, for example, Hua \textit{et al.}~\cite{hua2018reverse} demonstrated the retrieval of network parameters, such as the number of layers, size of filters, data dependencies among layers, etc. Then, Batina \textit{et al.}~\cite{batina2019csi} proposed a full reverse engineering of neural network parameters based on power/EM side-channel analysis. The proposed attack is able to recover hyperparameters i.e., activation function, pre-trained weights, and the number of hidden layers and neurons in each layer, without access to any training data. The adversary uses a combination of simple power/EM analysis, differential power/EM analysis and timing analysis to recover different parameters. Yu \textit{et al.}~\cite{yu2020deepem} proposed a model extraction attack based on a combination of EM side-channel measurement and adversarial active learning to recover the Binarized Neural Networks (BNNs) architecture on popular large-scale NN hardware accelerators.
The network architecture is first inferred through EM leakage, and then, the parameters are estimated with adversarial learning. As such, it can be observed that SCA has been widely adopted for the attack against neural networks.

\subsection{Countermeasures for Hardware Attacks on Neural Networks}
The first attempt to thwart side-channel analysis of neural networks was called MaskedNet~\cite{dubey2020maskednet}.
As indicated by the name, the authors utilize the masking countermeasure to partially protect the networks.
They developed novel hardware components such as masked adder trees for fully connected layers and masked ReLUs.
The approach works with binarized neural networks which use binary weights and activation values.

The same authors later extended their approach to fully protect neural networks by boolean masking~\cite{dubey2020bomanet,dubey2022guarding}.
While the latency overhead is relatively small (3.5\%), the area overhead is noticeable (5.9$\times$).
The overheads were further reduced in~\cite{dubey2022modulonet} where the authors used domain-oriented masking.

A threshold implementation (TI) with 64\% area and 5.5$\times$ energy overhead was presented in~\cite{maji2022threshold}.
For generating random numbers required for the TI, Trivium stream cipher was used.

Apart from masking-related approaches, garbled circuits were used to protect not only against SCA, but to enhance the privacy of the underlying models~\cite{hashemi2022hwgn2}.

To the best of our knowledge, the only software-based approach for protecting the neural networks against SCA uses shuffling~\cite{nozaki2021shuffling}.
The authors randomize the order of execution for multiplications within neurons.
This, however, does not prevent timing attacks on determining the activation function.

\section{Timing Analysis of Activation Functions}
In this section, we first explain the experimental setup used for our measurements.
Then, we provide the timing analysis without the countermeasures to better understand the timing pattern.

\subsection{Experimental Setup}
\label{sec:setup}
The experiments in this \paper were done by using a NewAE ChipWhisperer tool\footnote{\url{https://www.newae.com/chipwhisperer}}.
ChipWhisperer-LITE was used to perform the power analysis and a 32-bit ARM Cortex-M4 (STM32F3) mounted on a UFO board was used as a target.
The ADC sampling frequency was set to 29 MHz while the target frequency was set to 7 MHz.
A trigger signal was set to high before the activation function computation and immediately set to low afterward to precisely determine the timing.
The source code for the sample programs was written in C, the \texttt{<math.h>} library was used for mathematical functions.
To allow for more precise measurements, compiler optimizations were disabled.

\subsection{Timing Analysis without Countermeasures}
The timing behavior of activation functions was first examined in~\cite{batina2019csi}.
The authors measured the execution time of three functions: rectifier linear unit (ReLU), hyperbolic tangent (tanh), and sigmoid.
The definitions of those functions are as follows:
\[
ReLU(x)=\begin{cases}
0 & \text{if } x\leq0\\
x & \text{otherwise}
\end{cases},
\]
\[
sigmoid(x)=\frac{1}{1+e^{-x}},
\quad
tanh(x)=\frac{e^x-e^{-x}}{e^x+e^{-x}}. 
\]

\begin{figure}[tb]
    \centering
    \subfloat[ReLU\label{1a}]{%
       \includegraphics[width=0.48\linewidth]{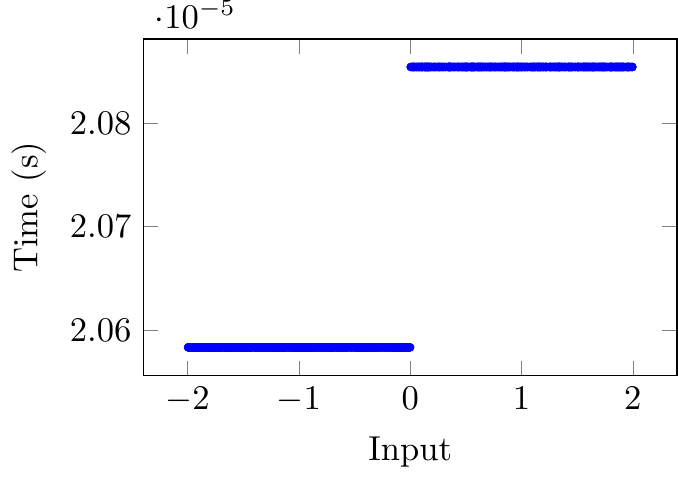}}
    \hfill
    \subfloat[sigmoid\label{1b}]{%
       \includegraphics[width=0.48\linewidth]{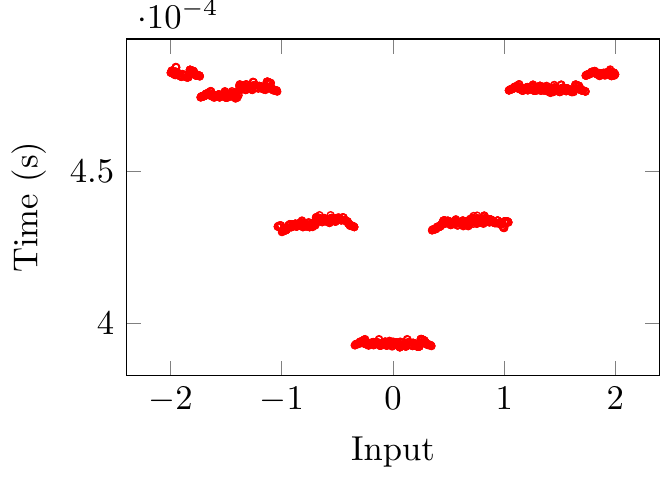}}
    \hfill
    \subfloat[tanh\label{1c}]{%
       \includegraphics[width=0.48\linewidth]{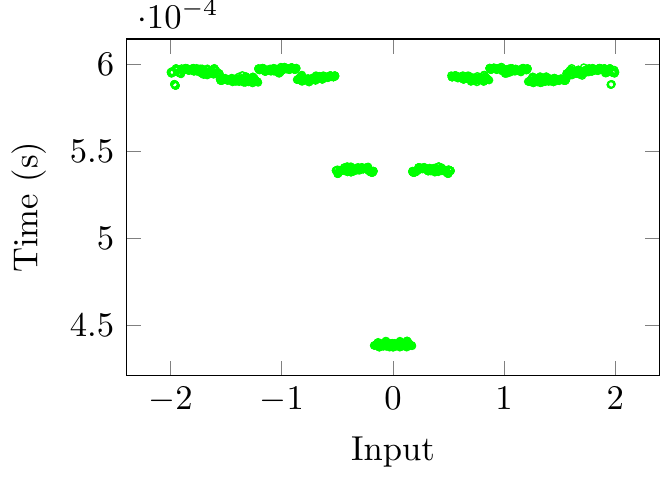}}
    \hfill
    \subfloat[relative comparison\label{1d}]{%
       \includegraphics[width=0.48\linewidth]{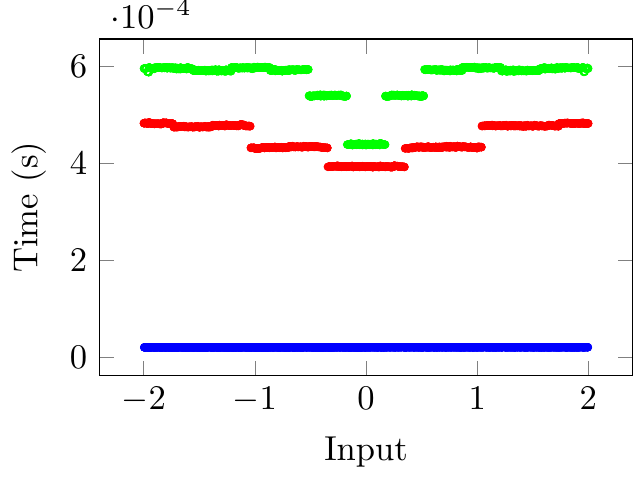}}
    \hfill
    \caption{Timing behavior of different activation functions. The patterns are clearly distinguishable if there is no side-channel protection in place.}
    \label{fig:original_functions}
\end{figure}

The conclusion was that each of these exhibits different timing patterns that are also dependent on the function input.
We repeated the experiment from~\cite{batina2019csi} and used $2000$ random inputs from $[-2,2]$ for the computations of each activation function.
We have obtained very similar results, see Figure~\ref{fig:original_functions} and Table~\ref{tab:time_original}.
This means that if the attacker sends a few queries and measures the time of the execution, they are able to distinguish which activation function is being used in a particular layer.

\begin{table}
    \centering
    \begin{tabular}{c|c|c|c}\hline
     Activation Function    &  Mean & Minimum & Maximum\\\hline
      Relu   & $0.0207$ & $0.0206$ & $0.0209$\\
      sigmoid & $0.4485$ & $0.3920$ & $0.4845$\\
      tanh &  $0.5170$ & $0.4375$ & $0.5985$\\\hline
    \end{tabular}
    \caption{Computation time (in milliseconds) for different activation functions.}
    \label{tab:time_original}
\end{table}

\section{Towards a Desynchronization-Based Countermeasure}
\label{sec:towards}
Desynchronization-based countermeasure was first introduced for cryptographic implementations~\cite{coron2009efficient,coron2010analysis,durvaux2013efficient}.
The main rationale of the desynchronization-based countermeasure is to remove the data dependency of the power consumption by randomizing the power consumption of the device during computation.

In this section, we will analyze the countermeasure when applied to activation function implementations.
For this purpose, we would like to utilize desynchronization to randomly delay the computation of each function such that it is impossible for the attacker to distinguish them from one another by examining the timing information. 

The computation times of different activation functions can be viewed as random variables depending on the inputs of the functions.
The time delay caused by additional desynchronization can be considered as another random variable $X$.
Since it is easy to generate a random variable with normal distribution in any programming language, we have decided to sample $X$ from a normal distribution.
A normal distribution is completely characterized by its mean and variance.
The mean specifies the expected average for the added delay and the variance characterizes how spread out the added delay is from the mean.
To randomize the computation, we would like to choose mean and variance for $X$ such that the resulting computation timing for all three functions follow a similar pattern.

To choose the mean, we look at the maximum and minimum possible timings for all three function computations.
Figure~\ref{fig:original_functions} shows that the computation times for each activation function are scattered into specific ``clusters'' depending on the input.
In particular, there are two, three, and three clusters for ReLU, sigmoid, and tanh respectively.
For example, when the input is positive, the computation time for ReLu is almost $2.09\times10^{-5}\ s$.
And when the input is near $0$, the computation time for tanh is around $4.4\times10^{-4}\ s$.
We have calculated the mean for the slowest cluster (i.e. the slowest cluster in tanh computation) which is $5.9\times 10^{-4}\ s$ and the mean for the fastest cluster (i.e. the faster cluster in ReLU computations) which is $2.06\times10^{-5}\ s$.
We have decided to choose the mean of $X$ to be $0.6$ milliseconds so that the very fast computations will have a chance to have comparable computation time as the slow ones.

Furthermore, to remove the distinct clusters in Figure~\ref{fig:original_functions}, we need to choose a variance big enough that the differences caused by the input values are negligible.
Table~\ref{tab:time_original} summarizes the data for computation times and we can see that the maximum is about $6\times 10^{-4}\ s$ and the minimum is about $2\times10^{-5}\ s$.
Thus we decided to choose the variance to have the same order of magnitude\footnote{The order of magnitude of a number $n$ is given by $b_n$ such that we can write $n$ in the form $n=a\times 10^{b_n}$, where $1/\sqrt{10}\leq a<\sqrt{10}$.} as $6\times 10^{-4}-2\times10^{-5}$, i.e. $0.1\times10^{-4}\ s^2$.

\begin{figure}
    \centering
    \includegraphics{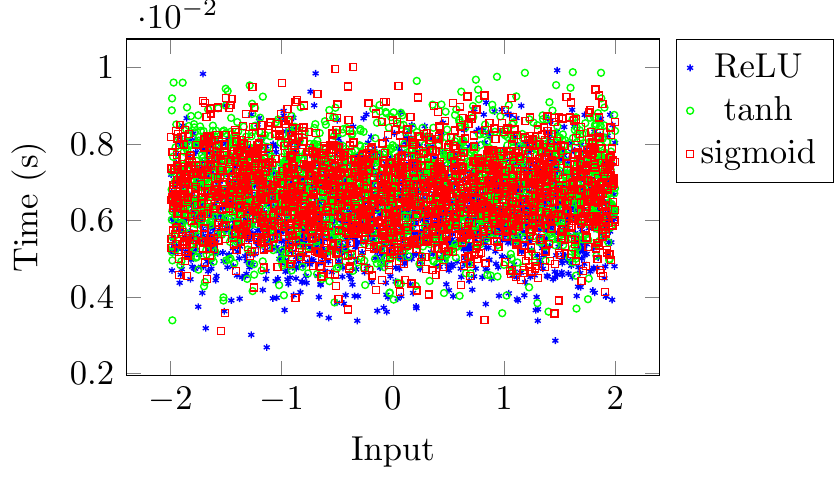}
    \caption{Timing behavior of activation functions with an applied desynchronization-based countermeasure. The timing patterns are not distinguishable.}
    \label{fig:protected}
\end{figure}

In summary, we add random desynchronization, whose delayed computation time is a sample from a normal distribution with mean $6\times 10^{-4}\ s$ and variance $0.1\times10^{-4}\ s^2$, to all three activation functions.
$2000$ random inputs are given to each of the functions.
The resulting computation times are shown in Figure~\ref{fig:protected} and Table~\ref{tab:time_jitter}.

\begin{table}
    \centering
    \begin{tabular}{c|c|c|c}\hline
     Activation Function    &  Mean & Minimum & Maximum\\\hline
      Relu   & $6.31$ & $2.69$ & $9.93$\\
      sigmoid & $6.72$ & $3.11$ & $10.01$\\
      tanh &  $6.81$ & $3.40$ & $9.88$\\\hline
    \end{tabular}
    \caption{Computation time (in milliseconds) for different activation functions with random desynchronization.}
    \label{tab:time_jitter}
\end{table}

In general, we propose the following steps for choosing a desynchronization-based countermeasure for protecting the computation of activation functions:
\begin{enumerate}
    \item Collect computation time data for all three activation functions.
    \item Compute the average of the fastest cluster of timings, denoted $t_f$, and the average of the slowest cluster denoted $t_s$.
    Let $\mu=t_f-t_s$.
    \item Compute the difference between the longest computation time and the shortest computation time, say $\Delta t$ seconds.
    Let $d_{\Delta t}$ denote the order of magnitude of $\Delta t$.
    Let $\sigma^2=1\times 10^{-d_{\Delta_t}}\ s^2$.
    \item Add random desynchronizations, whose delayed computation time is a sample from a normal distribution with mean $\mu$ and variance $\sigma^2$, to the implementations of the activation functions.
\end{enumerate}

\section{Leakage Assessment of a Neuron Computation}

\begin{figure}[tb]
    \centering
    \begin{tabular}{l}
         \includegraphics{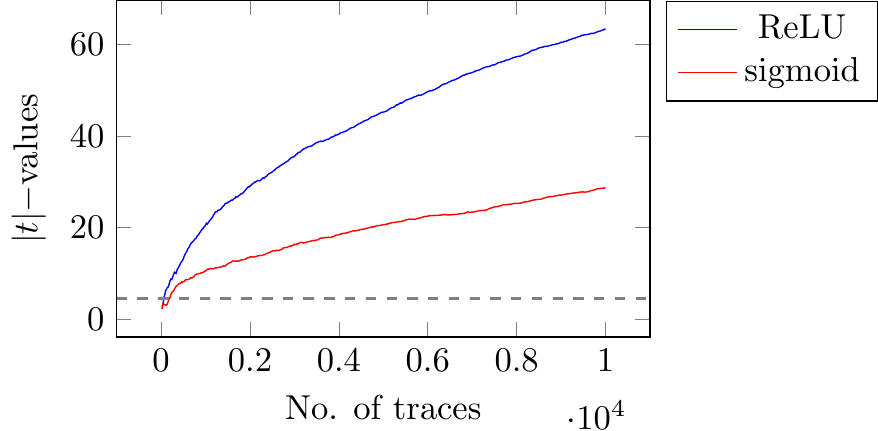}\\
    \hspace{1.5mm}\includegraphics{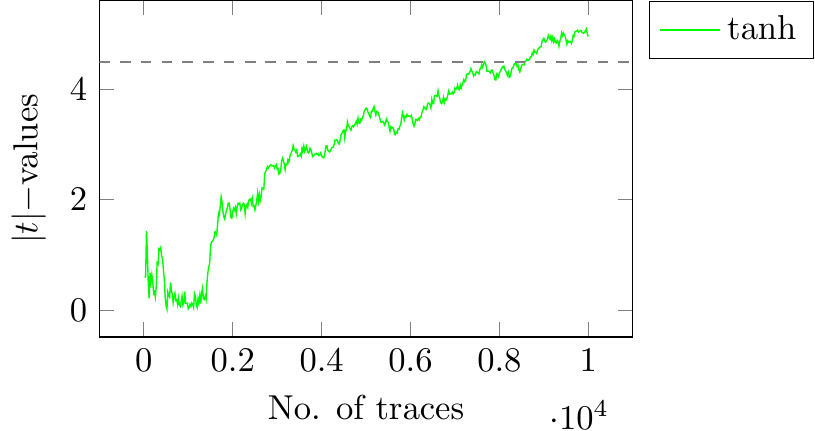}
    \end{tabular}
    \caption{TVLA results for ReLU, sigmoid, and tanh without the application of the countermeasure.}
    \label{fig:tvla_nojitter}
\end{figure}

\begin{figure}
    \centering
    \includegraphics{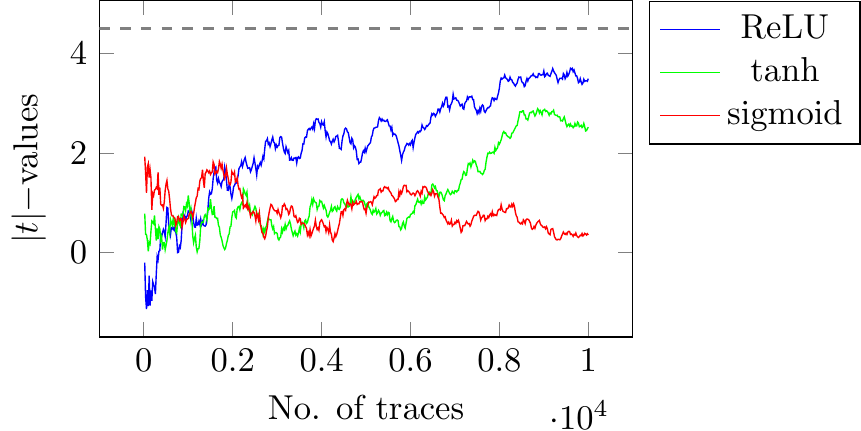}
    \caption{TVLA results for ReLU, sigmoid, and tanh with the application of the desynchronization-based countermeasure.}
    \label{fig:tvla_jitter}
\end{figure}

In order to evaluate the performance, we used Test Vector Leakage Assessment (TVLA)~\cite{Goodwill2011ATM}. 
The idea was to perform a t-test on a dataset from fixed vs random inputs, and in our case, on measured timing from the inference execution.

Here, we detail the computations for ReLU.
Those for tanh and sigmoid are done similarly.
First, we compute the execution time of when the neural network~\footnote{We are using a similar model architecture as~\cite{batina2019csi} for MNIST dataset as a proof of concept.} is running the inference for a fixed input (which is chosen randomly at the beginning but fixed for the rest of experiment). We measured the timing of $5000$ inference executions.
Let us denote those timings by $x_1,x_2,\dots,x_{5000}$.
Then, similarly, we calculate the execution times for given random inputs, where each is chosen randomly and different from each inference execution. 
Let us denote the corresponding timings by $y_1,y_2,\dots,y_{5000}$.
By the TVLA method, we compute
\[
t=\frac{\overline{x}-\overline{y}}{\sqrt{\frac{\sigma_x^2}{5000}+\frac{\sigma_y^2}{5000}}},
\]
where $\overline{x}$, $\overline{y}$, $\sigma_x^2$ and $\sigma_y^2$ are the mean of $x_i$, mean of $y_i$, variance of $x_i$ and variance of $y_i$ respectively.
In case the absolute values of $t$, called $|t|$-values, cross the threshold of 4.5, it can be concluded there is a data-dependent leakage in the measured traces.
As can be seen in Figure~\ref{fig:tvla_nojitter}, the TVLA test for the timing of the activation functions shows leakage, as expected.
On the other hand, after the application of the proposed countermeasure, $|t|$-values stay below the threshold, as can be seen in Figure~\ref{fig:tvla_jitter}.

\section{Discussion}
\subsection{Overheads} 
While it might seem from the figures that the timing overhead of the countermeasure is relatively high, it is to be noted that the activation function is only a small part of the entire neural network computation.
The majority of the computation is spent on multiplications that are dependent on the number of neurons in the layers.
For example, using the device from Section~\ref{sec:setup}, the time for computation of one multiplication is roughly $1.165\times 10^{-5}$ seconds and that for addition is about $1.124\times10^{-5}$ seconds.
The computation timing of the activation function is roughly $0.21 - 5.99\times10^{-4}$ seconds without desynchronization (Table~\ref{tab:time_original}) and $3.11 - 10.01\times10^{-3}$ seconds with random desynchronization (Table~\ref{tab:time_jitter}).
In modern architectures, there are thousands of multiplications and additions with just one activation function computation for one neuron.
VGG-19~\cite{simonyan2014very}, one of the popular public networks for ImageNet classification challenge, has $4096$ neurons in the last hidden layer and $1000$ neurons for the output layer, which amounts to $4096$ multiplications and $4096$ additions for each output neuron.
In this case, the computation time for multiplications and additions in one neuron is about $0.09$ seconds, and that for the whole neuron computation is $0.09002 - 0.0906$ seconds without desynchronization or $0.093 - 0.1$ seconds with the desynchronization countermeasure.
Thus, the overhead for the computation of one neuron is between $2.6\% - 11\%$.
We would also like to note that this is purely the timing for the calculations, not taking into account the memory operations -- if the entire computation is considered, the overhead would be even lower. As this is a proof-of-concept and still a work in progress, we observe a positive trend and will be further investigating this in future works.

\subsection{Other Activation Functions}
Batina, \textit{et al.}~\cite{batina2019csi} show results on softmax activation function apart from the three functions that were analyzed in this work.
The softmax function is normally used in the output layer of a neural network to transform the raw outputs (logits) into a vector of probabilities.
As it is unusual to find softmax in other layers of the network, we did not consider this activation function in our work.

There are other activation functions that can be used, for example leaky ReLU, exponential linear unit (ELU), etc.
We argue that the process of applying the desynchronization-based countermeasure on these functions is the same as described in Section~\ref{sec:towards}.



\section{Conclusion}

SCA has been a threat for neural networks model extraction, as it could perform reverse engineering to reconstruct the secret parameters of the networks. One of the main critical component of the network is the activation function, which as shown in previous works, is vulnerable against timing attack.
In this work, we have investigated desynchronization-based countermeasures, from SCA domain to hide the timing leakage behavior. Our experimental results then shown that desynchronization based approach can succesfully hide the timing leakage information which could be used to prevent model extraction attack of the activation parameter.
The overhead ultimately depends on the number of neurons in the fully-connected layer, for example, in the case of 4096 neurons in VGG-19, the overheads are between 2.8\% and 11\%.

\subsection*{Acknowledgement}
This work has been supported in parts by the ``University SAL Labs'' initiative of Silicon Austria Labs (SAL) and its Austrian partner
universities for applied fundamental research for electronic based systems.
This project has received funding from the European Union's Horizon 2020 Research and Innovation Programme under the Programme SASPRO 2 COFUND Marie Sklodowska-Curie grant agreement No. 945478.

\bibliographystyle{IEEEtran}
\bibliography{bibl}






\vfill

\end{document}